# Rebuttal to Schmelzer and Tropin, "Glass Transition, Crystallization of Glass-Forming Melts, and Entropy" [*Entropy* 20 [2] 103 (2018)]

Edgar D. Zanotto[1,*] and John C. Mauro[2,*]

[1]Department of Materials Engineering, Center for Research, Technology and Education in Vitreous Materials, Federal University of São Carlos, SP, Brazil

[2]Department of Materials Science and Engineering, The Pennsylvania State University, University Park, Pennsylvania, USA

* Corresponding authors. E-mail: dedz@ufscar.br, jcm426@psu.edu

**Abstract:** In a recent article, Schmelzer and Tropin [*Entropy* 20 [2] 103 (2018)] presented an unfounded, confusing critique of several aspects of modern glass science. Relying on pre-Socratic Greek philosophy and state-of-the-art scientific understanding from the 1920s-1930s, Schmelzer and Tropin propagate an antiquated view of glass physics that is at odds with well-accepted knowledge in the field from both theory and experiments conducted in the post-World War II era. The objective of this short letter is to elucidate and extinguish their critique. This rebuttal is directed to our colleagues and especially to students who might otherwise become confused.

**I. Introduction**

This short letter serves as a rebuttal to a recent article by Schmelzer and Tropin [1], which was published in the MDPI journal *Entropy* despite two strong rejections. This is not meant to be a comprehensive listing of everything that is inaccurate in their article. Rather, it is meant to address some of their main points attacking modern advances in glass science in favor of the reactionary views from the early 20th century. Here we adopt the format of quoting relevant passages from Schmelzer and Tropin, followed by suitable rebuttals to those sections.

**II. Definition of "Glass"**

Much of the content of Schmelzer and Tropin's paper is devoted to an attack on our recently proposed modern definitions of "glass" [2]. The first definition presented in our paper is meant for the general public and non-experts in the field: "*Glass is a nonequilibrium, non-crystalline state of matter that appears solid on a short time scale but continuously relaxes towards the liquid state.*" We also proposed an alternative, more detailed definition for use by advanced students and professionals in the field: "*Glass is a nonequilibrium, non-crystalline condensed state of matter that exhibits a glass transition. The structure of glasses is similar to that of their parent supercooled liquids (SCL), and they spontaneously relax toward the SCL state. Their ultimate fate is to solidify, i.e., crystallize.*" According to a recent Mendeley report, our article, Ref. [2], was viewed by approximately 6,000 researchers in the eight months since publication.

While our definition stresses the hybrid nature of the glassy state, combining both solid-like and liquid-like features, Schmelzer and Tropin insist upon older definitions from the 1920s-1930s, where glass is treated as strictly a solid. In their Section 2.1, "Basic definitions and some comments," Schmelzer and Tropin write:

"As described in detail in [24] (of their manuscript) a first definition of glasses was proposed by Gustav Tammann (see, e.g., his monograph [40]), denoting it as undercooled solidified melts. This definition was considerably expanded by Simon [30,31,32], who suggested to consider glasses as kinetically frozen-in thermodynamically non-equilibrium systems, distinguishing glasses from amorphous systems in thermodynamic equilibrium. Hereby it is assumed in a first approximation—suggested also by Simon—that the transformation takes place at some well-defined discrete temperature, the glass transition temperature, $T_g$. Simon was aware, of course, that the glass transition proceeds not at a discrete temperature value but over a certain temperature range. He had already also noted with reference to Tammann and Kohlhaas [41] and Parks and Huffmann [42] explicitly that by varying the cooling rate, different glasses can be obtained. However, he considered both the width of this glass transition range and the effect of varying cooling rates on glass properties as small and, for this reason, to be of minor importance [32]. Such reservations are today known to be not adequate, particularly if wide ranges of cooling and heating rates are employed while available [43].

Since these first fundamental considerations by Tammann and Simon, a variety of different approaches has been advanced in order to understand the detailed mechanism of the glass transition and, in its connection, the nature of the vitreous state (e.g., [44,45,46,47,48]; an overview is given in [49]). However, despite the different treatments, when posing the question as to which of the principal states of matter (solid, liquid, or gas; see, e.g., [50]) glass belongs to, in line with the classical treatment of Simon and Tammann, **glasses have to be considered as a solid**. Glasses behave as a solid in the absolute majority of applications. Indeed, as noted in a frequently cited statement by the Nobel laureate P. W. Anderson, "*The deepest and most interesting unsolved problem in solid state theory is probably the theory of the nature of glass and the glass transition*" [51]."

These classical papers of Simon (1927) and Tammann (1933) are also cited in our recent article [2], but they do not change our two proposed definitions for glass, which are based on modern understanding of glass structure and thermodynamics. As we explain in Ref. [2], the glass transition temperature ($T_g$) is observed when the characteristic structural relaxation time of a supercooled liquid (SCL) is similar to the experimental observation time. In other words, $T_g$ happens when the Deborah number is 1. Below $T_g$, the structure of the SCL is (only temporarily) frozen in the vitreous state.

It is also quite surprising that Schmelzer defends these very old, incomplete definitions of glass, but he has proposed his own unwieldy definition in 2013 [3]: "*Glasses are thermodynamically non-equilibrium kinetically stabilized amorphous **solids**, in which the molecular disorder and the thermodynamic properties corresponding to the state of the respective under-cooled melt at a temperature $T^*$ are frozen-in. Hereby $T^*$ differs from the actual temperature $T$.*"

**II. On Greek Philosophy**

Schmelzer and Tropin [1] continue their argument by drawing upon the pre-Socratic Greek philosopher, Heraclitus:

> "Already by this reason we **consider it as unreasonable** to treat glasses as "*a state of matter that appears solid on a short time scale but continuously **relaxes** towards the liquid state*" as supposed in an advanced recent modification of the definition of glass by Zanotto and Mauro [52]. The latter property—to change its state in time scales "*which exceed the limits of human history*" (see discussion below)—is not a specific feature of glasses as was well-known already by **Heraclitus**. His "*pantha rhei*" or "*everything flows*" refers not only to glasses. **Crystals, rivers, mountains** (see the subsequent discussion of the Deborah number), and so forth also flow on sufficiently large time scales. Because **everything flows** on such historical time scales, this feature is not a specific property of glasses and cannot be used to distinguish it from any other states of matter."

"Heraclitus" is invoked five times in the Schmelzer-Tropin article, despite the fact that this has nothing to do with glass. While Heraclitus certainly observed (liquid) rivers flowing in his time scale, we wonder how Heraclitus has measured the flow of crystals during his lifetime, 535 – 475 BC? Where are his experimental results published?

**IV. Glass Flow vs. Relaxation**

Schmelzer and Tropin [1] then fast-forward to the 1850s-1920s, writing:

> "The fact that predominantly glasses **flow** with a perceptible rate on relevant time scales only in a certain temperature range and not beyond is well known in glass technology [24], as is clearly formulated also for example by Tammann [40,41] and is already given in the title of his well-known paper by Tool [53]. Of course, for certain applications, flow processes have to be taken into consideration as is well known already from the work of R. & F. Kohlrausch, Weber, Williams & Watts, Adams & Williamson, Eyring & Tobolsky (see, e.g., [54]) starting around 1850. However, such possible flow processes under certain conditions have not been considered as essential by Tammann and Simon in their definition of glass. Such a point of view, that flow processes may be neglected in most applications for relevant times scales, has also been clearly expressed by one of the authors of [52] in [55,56]. For example, in [55] it is noted that "*window glasses may flow at ambient temperature only over incredibly long times, which exceed the limits of human history*". The flow processes considered in [55,56] are primarily the response to external fields and are governed by viscosity. However, the viscosity and structural relaxation time are uniquely correlated as noted also in [55,56], where the analysis is performed widely in terms of relaxation times."

References [55] and [56] in [1] (Refs. [4]-[5] in this letter) are indeed authored by one of us (Zanotto). In these two articles, we estimated the average *relaxation* times (in the absence of any external stress) – *not* the flow times under stress – of $GeO_2$ glass and a commercial soda-lime-silica glass at room temperature, i.e., approximately 530ºC below their laboratory glass transition temperatures. The calculated relaxation times are indeed astronomical, but they are still **finite!**

More recently, Mauro and co-authors [6] used an improved model and new experimental viscosity measurements to calculate the times needed for a real medieval glass – the Westminster Abbey cathedral glass – to relax (under zero external stress) and also to flow under gravity. The results are billions of years for both times. But the conclusion is that glass will indeed flow at room temperature: one just needs to find a persevering student that is willing to wait that long for his experiment!

Also, Schmelzer and Tropin [1] misunderstood our definition and confused flow with relaxation! Our shorter definition reads [2]: *"Glass is a state of matter that appears solid on a short time scale but continuously relaxes towards the liquid state."* Yes, that is exactly what we meant to say: "relaxes." Glasses are thermodynamically unstable against the supercooled liquid (SCL) and spontaneously relax – even in the absence of external stress – towards the SCL state. Solids do not undergo spontaneous relaxation. They may creep or flow (by different mechanisms than glasses) under external stress.

Schmelzer and Tropin continue [1]:

"In [52], Zanotto and Mauro discuss Simon's definition of glass as a freezing-in process. Referring to [30] and presenting Simon's point of view in the form that "*glass is a rigid material obtained from freezing-in a supercooled liquid in a narrow temperature range*", they further state that "*it is not clear if he intended to convey the same meaning we are using here (frozen = a temporary state)*". However, the meaning Simon assigned to his statement of freezing-in is clearly reflected in [32]. In free translation, it sounds as though freezing-in at $T_g$ does not imply that below $T_g$ relaxation processes are excluded. However, any such structural transformations proceed already slightly below $T_g$ with such large time scales that the **suggestion of a permanent arrest of such structural changes is completely substantiated** ([32], p. 223); or, as stated by Davies and Jones ([57], p. 375), "*Simon pointed out that as a glass is cooled through *its transformation temperature the molecular diffusion which is necessary to effect the appropriate change in configuration is increasingly inhibited and finally becomes practically impossible*". This interpretation is fully in line with [55,56] but not with the revised definition of a glass given in [52]."

The relaxation times of a glass strongly depend upon the chemical composition (e.g., the liquid fragility), the temperature (i.e., how far below $T_g$ the material is), and the entire thermal history (i.e., the degree of thermal disequilibrium). For small undercoolings below $T_g$, the relaxation times (and flow under stress) will be only minutes or hours. However, if one is studying his/her glass well below the $T_g$, the relaxation and flow times will be extremely long, as shown in [4]-[9]. But, in the end, they all relax and flow at any temperature above the absolute zero! This is a requirement for a nonequilibrium, thermodynamically unstable system with finite thermal energy.

**V. Viscosity**

Schmelzer and Tropin [1] draw upon the 1920s notion of divergent viscosity in glass-forming liquids, writing:

"Moreover, the viscosity of glass-forming melts increases dramatically with decreasing temperature [24,58]. **One of the relations** describing it with a sufficiently high degree of accuracy for most applications is the Vogel–Fulcher–Tammann equation widely employed in glass science [59]. This equation results in a **divergence of the viscosity** at finite values of temperature, denoted as Vogel temperatures. Whether the viscosity will really diverge or not is a matter of intensive debate; it cannot be established by direct experimental investigations restricted to maximum values of viscosity less than $10^{18}$ Pa·s. In any case, a variety of models of the vitreous state lead to the confirmation of such a conclusion. **However once the viscosity diverges**, the structural relaxation time also diverges. Glasses at temperatures below the Vogel temperature are then excluded from the vitreous state by the above-mentioned definition."

This passage calls for a reality check on the Vogel-Fulcher-Tammann (VFT) equation, which was proposed in the 1920s as an empirical function to describe the viscosity of supercooled liquids [10]. Here, Schmelzer and Tropin make several mistakes. First, the VFT equation does not describe the viscosity of the nonequilibrium glassy state, which is typically many orders of magnitude less than that of the corresponding supercooled liquid at the same temperature. This is well known, both through experimental measurements and basic theory [7], [9]. Second, the divergence predicted by the VFT equation (for supercooled liquids, not glasses!) is an artifact of the particular form of this equation. Again, VFT is an empirical equation, not a physical law! Since the 1920s, several more successful and physically derived models have been proposed that do *not* predict such divergence and have been shown to provide dramatically improved the description of liquid viscosity at low temperatures. Unfortunately, none of this work is cited by Schmelzer and Tropin, since it does not fit with their agenda of promoting reactionary scientific viewpoints. Readers interested in a review of the modern understanding of the viscosity of glass-forming systems are referred to Ref. [9].

**VI. Crystallization**

Schmelzer and Tropin [1] also reject the reality of glass crystallization, writing:

"Further extending **their modification of the definition of glass**, Zanotto and Mauro propose to include into the definition of glass the statement, "*Their ultimate fate, in the limit of infinite time, is to **crystallize***". However, even if this statement would be true, it seems to us not to be reasonable to include such a statement into the definition, as it does not supply any additional information as to what glasses are. In addition, if at all, crystallization proceeds at a perceptible rate for states below the glass transition range also only at time scales exceeding the limits of human history."

Here, Schmelzer and Tropin continue to confuse the reader because they show only a part of our extended definition of glass, which reads [2]: "*Glass is a nonequilibrium, noncrystalline condensed state of matter that exhibits a glass transition. The structure of glasses is similar to that of their parent supercooled liquids (SCL), and they spontaneously relax toward the SCL state. Their ultimate fate is to solidify, i.e., crystallize.*"

We stress here that this definition is for advanced students who understand the meaning of glass transition. Glasses will always relax and then crystallize! Moreover, in certain cases, they can even crystallize before full relaxation [2]. As for relaxation, the actual crystallization time depends upon the chemical composition of the material and the temperature. If one is testing a glass a few degrees below its $T_g$, these processes will take only minutes or hours; if one is working well below $T_g$, say at room temperature for a typical oxide glass, crystallization will take longer than the age of the universe. To give a few examples, in the Zanotto laboratory we have crystallized lithium diborate glass and a diopside (CaO·MgO·2SiO$_2$) glass, 30-50 °C below their respective $T_g$ in less than 3 months [11]. Crystallization (or "devitrification") is also a well-known critical problem for the glass industry [12]-[13].

As shown in Ref. [12], even the most stable oxide glasses known so far, B$_2$O$_3$ and albite, will crystallize below their $T_g$ with adequate heat treatments. The crystallization times are indeed very long well below $T_g$, but they are **finite**!

In our opinion, this part of the definition is very important because it reinforces the fact that glasses are not true solids: they only solidify at the end of their lives, when they crystallize! Our message is that all of the dynamic processes that take place above the laboratory $T_g$ (ionic diffusion, viscous flow, relaxation, liquid-liquid phase separation, crystal nucleation and growth, etc.) also take place below the $T_g$. It only takes longer and longer as the temperature drops.

**VII. Broken Ergodicity and Entropy**

Schmelzer and Tropin are also profoundly confused regarding the concepts of broken ergodicity and the configurational entropy of glass. They write [1]:

> "Our conclusion is that the treatment of vitrification as a process of continuously breaking ergodicity with entropy loss and a residual entropy tending to zero in the limit of zero absolute temperature is in disagreement with the absolute majority of experimental and theoretical investigations of this process and the nature of the vitreous state. This conclusion is illustrated by model computations."

In fact, their model is constructed to give the results that confirm their preconceived biases regarding glass entropy. They do not account for the reality that broken ergodicity (i.e., the non-equivalence of time and ensemble averages of properties due to the long timescales involved with glass science) is the single most fundamental defining feature of the glassy state [14]-[16]. One cannot accurately model the thermodynamics or statistical mechanics of the glassy state without accounting for non-ergodic nature. This was demonstrated clearly by Mauro et al. [17] who derived model-free equations for the entropy of glass considering both ensemble and time averages. A finite residual entropy is present only in the ensemble-average formulation, which would also predict a non-zero heat capacity of glass at absolutely zero temperature. In contrast, the time-average formulation yields zero entropy at absolute zero, since the glass is confined to a single microstate, and all of its properties are an average over this one microstate alone. Without fluctuations, the time-average formulation yields zero heat capacity in the limit of absolute zero temperature, which is consistent with experimental results [17]. While it is possible to construct a model to obtain any result that one desires, as Schmelzer and Tropin have done [1], the ultimate truth must come from experiments, where there is no doubt that glasses are confined to a single microstate in the limit of absolute zero, and hence have zero heat capacity and zero entropy, consistent with the Third Law of Thermodynamics [17].

**VIII. Conclusions**

We have given strong arguments, based on recent literature, to show that the use of Greek philosophy and ideas of the 1920s may lead to a gross misunderstanding of key phenomena regarding the glassy state: viscous flow, relaxation, crystallization, and entropy of supercooled liquids and glasses. And this was indeed the principal motivation to write our recent article, Ref. [2].

Science is about progress. Scientists in each generation build upon the great discoveries of those in prior generations. As time progresses, more sophisticated theories and experiments can reveal new understanding about the infinitely complex universe in which we inhabit. Progress is something that should be fully embraced by all researchers in the scientific community. In the case of glass science, dramatic advances have been made in understanding the fundamental physics and chemistry of glass and applications to advanced glass technology and new product design. The development of new glass products directly helps to improve the quality of life for human civilization follows. But these new products can only be designed through an improved fundamental understanding of the materials processes driving their properties and behavior [18]. It is therefore essential to embrace progress and reject the regressive views of those who would wish to set the clocks of our understanding back by nearly a century.